\documentclass[submission,copyright,creativecommons]{eptcs}
\providecommand{\event}{ICLP 2026} 

\usepackage{iftex}

\ifpdf
  \usepackage{underscore}         
  \usepackage[T1]{fontenc}        
\else
  \usepackage{breakurl}           
\fi

\usepackage{todonotes}

\usepackage{amsmath}
\usepackage{algorithmicx}
\usepackage{caption}
\usepackage{bsymb}
\usepackage{wrapfig}
\usepackage{tabularx}
\usepackage{booktabs}
\usepackage{mathpartir}
\usepackage{tikz}
\usepackage{makecell}
\usetikzlibrary{svg.path,calc,shapes.geometric}
\usepackage{xcolor}
\definecolor{prob2uicolor}{HTML}{037875}
\definecolor{lightblue}{RGB}{173,216,230}
\definecolor{orcidlogocol}{HTML}{A6CE39}
\tikzset{
	orcidlogo/.pic={
		\fill[orcidlogocol] svg{M256,128c0,70.7-57.3,128-128,128C57.3,256,0,198.7,0,128C0,57.3,57.3,0,128,0C198.7,0,256,57.3,256,128z};
		\fill[white] svg{M86.3,186.2H70.9V79.1h15.4v48.4V186.2z}
		svg{M108.9,79.1h41.6c39.6,0,57,28.3,57,53.6c0,27.5-21.5,53.6-56.8,53.6h-41.8V79.1z M124.3,172.4h24.5c34.9,0,42.9-26.5,42.9-39.7c0-21.5-13.7-39.7-43.7-39.7h-23.7V172.4z}
		svg{M88.7,56.8c0,5.5-4.5,10.1-10.1,10.1c-5.6,0-10.1-4.6-10.1-10.1c0-5.6,4.5-10.1,10.1-10.1C84.2,46.7,88.7,51.3,88.7,56.8z};
	}
}
\newcommand{\orcid}[1]{%
	\resizebox{8px}{8px}{
		\href{https://orcid.org/#1}{\tikz[yscale=-1,transform shape]{\pic{orcidlogo}}}}%
}
\usepackage{diagbox}

\DeclareUnicodeCharacter{2208}{$\in$}
\DeclareUnicodeCharacter{2209}{$\notin$}
\DeclareUnicodeCharacter{2227}{$\land$}

\newcommand{\prob}{\textsc{ProB}}
\newcommand{\probtwoui}{\textsc{ProB2-UI}}

\newcommand{\visb}{\textsc{VisB}}
\newcommand{\simb}{\textsc{SimB}}

\newcommand{\ignore}[1]{}

\usepackage{hyperref}
\usepackage[noabbrev]{cleveref}
\usepackage{color}

\usepackage[final,frozencache,cachedir=minted-cache]{minted2}

\DeclareCaptionType{mintlisting}[Listing][List of Listings]
\counterwithout{mintlisting}{section}
\crefname{mintlisting}{Listing}{Listings}
\Crefname{mintlisting}{Listing}{Listings}

\title{
	    Animation, Verification and Visualisation of Prolog Transition Systems with \prob{}
         }
\author{Jan Gruteser \orcid{0009-0006-4228-404X} \qquad Michael Leuschel \orcid{0000-0002-4595-1518} \qquad Katharina Engels \orcid{0009-0001-3595-3683} \qquad Fabian Vu \orcid{0000-0003-2556-5553}
\institute{Faculty of Mathematics and Natural Science, Institute of Computer Science,\\
	Heinrich Heine University Düsseldorf, Universit\"{a}tsstr. 1, D-40225 D\"{u}sseldorf}
\email{\quad \{jan.gruteser,michael.leuschel,katharina.engels,fabian.vu\}@hhu.de}
}
\def\titlerunning{Animation, Verification and Visualisation of Prolog Transition Systems with \prob{}}
\def\authorrunning{J. Gruteser, M. Leuschel, K. Engels \& F. Vu}
\begin{document}
\maketitle

\begin{abstract}
	\prob{} is a Prolog-based model checker, animator and constraint solver for high-level formal specifications.
	One can also use \prob{} to animate transition systems defined by 
	Prolog predicates, allowing the application of its various 
	validation techniques.
    In this work, we present the existing features of \prob{}'s Prolog animation mode and its recent extensions.
    The extended capabilities 
    include simulation for statistical checks, more reliable trace replay, transitions with user input and improved state visualisation.
    We apply the new features to case studies, particularly for evaluating different strategies in game play, such as Connect Four.
    The features are useful for many other applications, especially for \prob{}'s new sequent prover for Event-B proof obligations, as well as for demonstration models for teaching in combination with interactive visualisation.
\end{abstract}

\section{Introduction}

\prob{}~\cite{DBLP:journals/sttt/LeuschelB08}\footnote{Our tooling IDE \probtwoui{} is available at \url{https://prob.hhu.de/w/index.php?title=Download\#ProB2-UI}.} is a model checker, animator, and constraint solver for high-level formal specifications implemented in SICStus Prolog.
While it primarily targets formal methods such as the B method~\cite{bmethod} and Event-B~\cite{Abrial10}, \prob{} can also load and animate transition systems specified in Prolog.
As a result, \prob{} supports most of its validation and verification features for Prolog, mainly animation, model checking and visualisation.
Moreover, one can use the interface to implement interpreters for other formalisms.
For historical reasons, the animation of Prolog systems is referred to as the XTL mode of \prob{}.\footnote{The name XTL originates from \emph{X}SB Prolog and \emph{T}emporal \textit{L}ogic. When first mentioned, XTL referred to a finite-state model checker for CTL properties of systems represented in XSB~\cite{farwer2004model,leuschel2002logic}. However, there is no longer a connection to XSB.}
This is how we refer to it in this article as well.

Motivated by several use cases, specifically \prob{}'s new sequent prover~\cite{sequent-prover-iclp}, the XTL mode has been subsequently enhanced by new features.
In the following, we first present the existing animation and model checking features, along with an introductory example.
Then, we discuss the new extensions, each accompanied by a motivating example. 
We apply the new integration with \simb{}~\cite{simb}, a timed-probabilistic simulator in \prob{}, to analyse a case study concerning game moves encoded as a Prolog transition system for Connect Four. 

Altogether, 
we present the following contributions:
\begin{itemize}
    \item new integration of \simb{} for simulation of Prolog models, allowing for statistical validation using Monte Carlo simulation, hypothesis tests, and estimators,
    \item improved state visualisation features with \visb{}~\cite{visb} and standalone HTML trace exports,
    \item new concept of symbolic transitions, enabling delayed computation to incorporate custom user input and to prevent premature execution of external side effects (e.g. a call to another tool),
    \item new static and dynamic transition annotations, useful for adding transition probabilities, textual descriptions and parameter names,
    \item support for more reliable and interactive trace replay,
    \item demonstration and evaluation of a case study in the context of game theory (Connect Four).
\end{itemize}

\noindent The source code for all discussed examples is available on GitHub:
\begin{center}
    \url{https://github.com/hhu-stups/xtl-examples}.
\end{center}
In the following section, we introduce the basic functionality of XTL.
\vspace{-0.4cm}

\section{Prolog-Based Transition Systems in \prob{}}\label{sec:background}

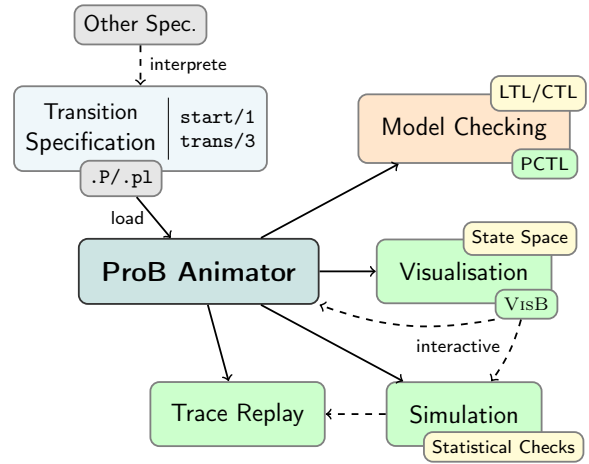
\begin{wrapfigure}[14]{r}{0.48\linewidth}
    \vspace{-0.4cm}
    \centering
    \resizebox{\linewidth}{!}{%
        \begin{tikzpicture}[auto, every loop/.style={},
            thick,
            node/.style={rectangle,font=\sffamily\normalsize},
            edge/.style={font=\sffamily\scriptsize}
            ]

            \node[node, rounded corners, draw=black!80, fill=prob2uicolor!20, inner sep=10pt] (animator) at (-1.6,-0.2) {\bf\sffamily\large \prob{} Animator};

            \node[node, rounded corners, draw=black!50, fill=orange!20, inner sep=10pt] (mc) at (2.5,2) {Model Checking};
            \node[node, rounded corners, draw=black!50, fill=yellow!20] (ltl) at ($(mc.north east) + (-0.5,0)$) {\scriptsize LTL/CTL};
            \node[node, rounded corners, draw=black!50, fill=green!20] (pctl) at ($(mc.south east) + (-0.4,0)$) {\scriptsize PCTL};

            \node[node, rounded corners, draw=black!50, fill=green!20, inner sep=10pt] (sim) at (2.5,-2.4) {Simulation};
            \node[node, rounded corners, draw=black!50, fill=yellow!20] (stat) at ($(sim.south east) + (-0.6,0)$) {\scriptsize Statistical Checks};
            
             \node[node, rounded corners, draw=black!50, fill=green!20, inner sep=9pt] (trace) at (-1.0,-2.4) {Trace Replay};

            \node[node, rounded corners, draw=black!50, fill=green!20, inner sep=10pt] (vis) at (2.5,-0.2) {Visualisation};
            \node[node, rounded corners, draw=black!50, fill=yellow!20] (statevis) at ($(vis.north east) + (-0.5,0)$) {\scriptsize State Space};
            \node[node, rounded corners, draw=black!50, fill=green!20] (visb) at ($(vis.south east) + (-0.4,0)$) {\scriptsize \visb};

            \node[node, rounded corners, draw=black!50, fill=lightblue!20, inner sep=5pt] (file) at (-2.5,2) {\makecell{\small Transition\\Specification}\hspace{.05cm} \vline\hspace{.05cm} \makecell{\footnotesize\texttt{start/1}\\[-.1cm]\footnotesize\texttt{trans/3}}};
            \node[node, rounded corners, draw=black!50, fill=gray!20] (P) at ($(file.south) - (0.3,0.1)$) {\footnotesize\texttt{.P/.pl}};

            \node[node, rounded corners, draw=black!50, fill=gray!20, inner sep=4pt] (other) at (-2.5,3.6) {\small Other Spec.};

            \draw[edge, ->] (animator) -- node[above] {} (mc);
            \draw[edge, ->] (P) -- node[left] {load} (animator);
            \draw[edge, dashed, ->] (other) -- node[right] {interprete} (file);
            \draw[edge, ->] (animator) -- node[right] {} (sim);
            \draw[edge, ->] (animator) -- node[right] {} (vis);
            \draw[edge, ->] (animator) -- node[above] {} (trace);
            \draw[edge, dashed, ->] (sim) -- node[above] {} (trace);
            \draw[edge, dashed, ->] (visb.south west) to[bend left=15] node[right] {} (animator.south east);
            \draw[edge, dashed, ->] (visb) to[bend left=10] node[left, yshift=1mm] {interactive} (sim);
        \end{tikzpicture}
    }
    \caption{\prob{}'s XTL Prolog Mode}
    \label{fig:overview}
\end{wrapfigure}

With \prob{}, it has long been possible to load Prolog files that define a labelled transition system by using special Prolog predicates.
Early versions of these transition predicates were introduced in \cite{leuschel1999infinite,leuschel2001design}.
The core of the definition is based on the predicates
\begin{center}
	\texttt{start(State)} and\\ \texttt{trans(Name,StateBefore,StateAfter)},
\end{center} which specify the \emph{initial states} and all possible \emph{transitions} to new states based on the current state.
A state can be represented by any (ground) Prolog term. 
For better comprehensibility of states and verification, additional \emph{state properties} can be provided by the predicate \texttt{prop(State,Prop)}.\footnote{More documentation can be found online at \url{https://prob.hhu.de/w/index.php?title=Other_languages}.}

By loading the transition system into the \prob{} animator, we obtain access to most of its validation and verification features, particularly animation and model checking.
An overview is shown in \Cref{fig:overview}.
The green nodes highlight features that have been added or improved in this work and will be described in the following sections.


\begin{mintlisting}[htb]
	\vspace{-.2cm}
	\captionof{listing}{\centering Simple XTL Prolog Specification of a Traffic Light System}
	\vspace{-.25cm}
	\hrule
	\vspace{-.1cm}
	\begin{minted}[linenos]{prolog}
 start(lights(red,red)).  % State: lights(Pedestrian,Cars)

 trans(tl_peds_green,lights(red,red),  lights(green,red)).
 trans(tl_peds_red,  lights(green,red),lights(red,red)).
 trans(switch_tl_cars(NL),lights(red,L),lights(red,NL)) :- tl_cars_seq(L,NL).

 prop(lights(P,_),'='(tl_peds,P)).
 prop(lights(_,C),'='(tl_cars,C)).
 prop(lights(P,C),unsafe) :- P \= red, C \= red.

 tl_cars_seq(red,red_yellow). tl_cars_seq(red_yellow,green).
 tl_cars_seq(green,yellow).   tl_cars_seq(yellow,red).
	\end{minted}
	\vspace{-.15cm}
	\hrule
	\vspace{-.9cm}
	\label{lst:traffic-light-example}
\end{mintlisting}

\paragraph{Animation.}


\begin{wrapfigure}[18]{r}{0.7\textwidth}
    \centering
    \includegraphics[width=\linewidth]{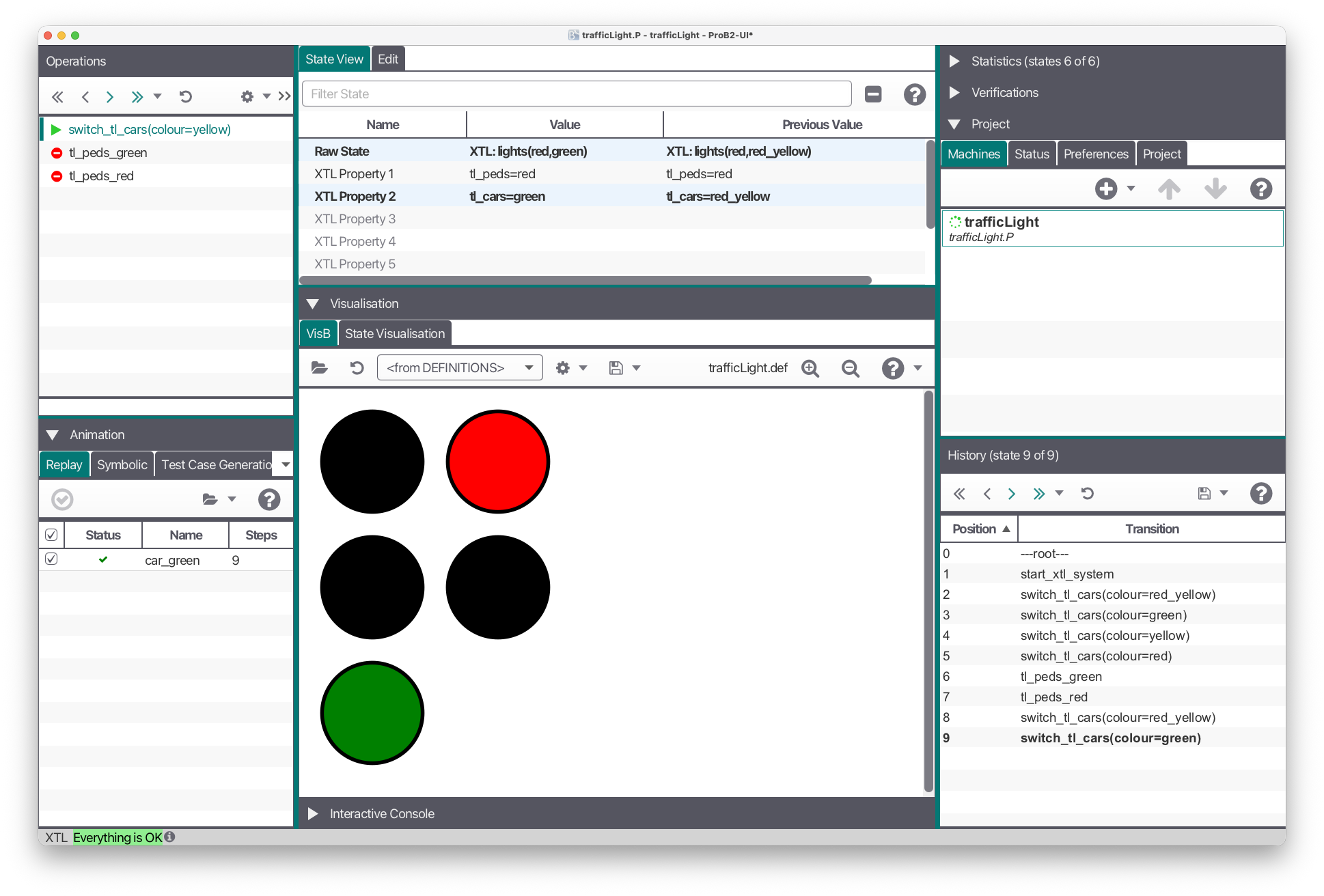}
    \vspace{-.7cm}
    \caption{Prolog Animation in \probtwoui{}}
    \label{fig:tlprob2ui}
\end{wrapfigure}
\Cref{lst:traffic-light-example} illustrates a simple example of a traffic light system, controlling one light for pedestrians and one for cars (available on GitHub~\cite{xtl-examples}).
The states are modelled as terms of the form \texttt{lights(P,C)}, where $P$ is the state of the pedestrian light and $C$ the state of the light for cars, given by Prolog atoms, e.g. \texttt{green} or \texttt{red\_yellow}.
Initially, both lights are \texttt{red} (line 1).
The transitions are specified in lines 3 to 4, which allow changing a traffic light if and only if the other one is \texttt{red}.
The first argument is the transition term, which can have additional parameters, such as \textit{NL} for \texttt{switch\_tl\_cars}.
Prolog predicates can be used freely in the computation of a transition; for example, \texttt{tl\_cars\_seq} (lines 11, 12) is used to define the sequence of colours.
If a transition predicate fails, no corresponding transition is available in the current state.

A user can open the specification file using one of the \prob{} interfaces: a command-line interface and graphical user interfaces based on Tcl/Tk and Java.
The latter is called \probtwoui{}~\cite{prob2-ui} and is shown in \Cref{fig:tlprob2ui} with the example of \Cref{lst:traffic-light-example} loaded into the animator.
By selecting one of the enabled transitions in the top-left corner, a user can animate a transition to the next state.
Here, only one transition is available for switching the car light to \texttt{yellow}.
The trace leading to the current state, i.e. a sequence of animated transitions starting at an initial state, is shown in the history at the bottom right.
Here, the current state is \texttt{lights(red,green)}, meaning the light for pedestrians is \texttt{red}, and the one for cars is \texttt{green}.
One can inspect the current state, with the specified state properties at the top and custom state visualisations at the bottom. 
We integrate the latter as part of this work and discuss it in \Cref{sec:visualisation}.

\paragraph{Model Checking.}

One can apply the \prob{} model checker to find deadlocks (states in which no transition is enabled) and \emph{unsafe} states. We define a state as unsafe if an invariant, that is, a safety property that must be satisfied in every state, is violated.
To declare safety conditions for the \prob{} model checker, one can define the special state property \texttt{unsafe}.
For example, line 9 of \Cref{lst:traffic-light-example} states that one of the two traffic lights must always be red.
For the traffic light system, the model checker explores six states and seven transitions without any unsafe state or deadlock (Fig.~\ref{fig:tlstatespace} shows a visualisation of the state space).

Additionally, \prob{} supports model checking of formulas in Linear Temporal Logic (LTL) and Computation Tree Logic (CTL) to verify temporal properties.
However, atomic propositions are restricted to state properties in the form of Prolog terms provided by \texttt{prop/2}.
For example, the following LTL formula~(left) verifies that \texttt{tl\_cars} will always ($G$ -- globally) eventually ($F$ -- finally) become \texttt{green} (which is prevented by the loop switching only the pedestrian light).
The CTL formula~(right) checks whether at least one such trace exists, which is indeed the case.
\begin{align*}
	GF \{\text{tl\_cars}=\text{green}\} \;\;\; \textit{\small(false)}\qquad \qquad EF \{\text{tl\_cars}=\text{green}\} \;\;\;\textit{\small(true)}
\end{align*}

To restrict the state space, we added support for \emph{scope} predicates in this work.
\prob{} will ignore all states where the predicate is not satisfied.
However, the predicate has to be provided in B syntax, so XTL properties are addressed by \prob{}'s built-in functions, for example, 
to obtain the current value of a state property (check that the car light does not show \texttt{red\_yellow}):
\begin{minted}{prolog}
     prob_pragma_string('SCOPE','STATE_PROPERTY("tl_cars") /= "red_yellow"').
\end{minted}
\texttt{prob\_pragma\_string/2} allows to keep preferences for \prob{} directly in the specification, where the preference name is the first and the value the second argument.
The predicate above eliminates all states after the car light has become \texttt{red\_yellow}, i.e. the model checker finds four states and four transitions.

There exist more advanced features, such as directed model checking controlled by a heuristic function~\cite{leuschel2010directed}, provided by \texttt{heuristic\_function\_result/2}.

\vspace{-.2cm}
\paragraph{Interpreters for other Formalisms.}

One can also use the interface to implement a custom \emph{interpreter} for other specification languages and to declare the transitions for \prob{}.
The interface was used, for instance, for Promela~(the verification language of the SPIN model checker~\cite{holzmann1997model}), SMV~\cite{mcmillan1993symbolic},\linebreak CSP(-M)~\cite{leuschel2001design},
and Lustre~\cite{vumasterthesis}.
Those interpreters are implemented to support the loading of files modelled in the new formalism.
From the engineering perspective, one must implement the formalism by defining predicates that describe the initial state (\texttt{start/1}), the state transitions involving predecessor and successor states (\texttt{trans/3}),
and state-based properties such as invariants and other state-based errors (\texttt{prop/2}).
Implementing those predicates enables the use of \prob{}'s animator and model checker for the new formalism, allowing one to check, e.g., for invariants and deadlocks.
Another advantage of implementing interpreters in Prolog is that operational semantics 
expressed as inference rules of the form [$\frac{\;\;B\;\;}{A} \; \textit{Cond}$] align with Prolog's clause notation and can be translated to \texttt{A~:-~B,~Cond}.
A concrete example is the following semantic rule, which one can encode as shown in \Cref{lst:interpreter}.\hspace*{3cm}
\begin{wrapfigure}[5]{r}{0.55\textwidth}
	\vspace{-.35cm}
	\captionof{listing}{\centering Interpreter Code for Semantic Rule}
	\vspace{-.25cm}
	\hrule
	\vspace{-.1cm}
	\begin{minted}{prolog}
 interpret(assign(x,E), State, NewState) :-
     eval(E,V),
     update(State, x, V, NewState).
    \end{minted}
	\vspace{-.15cm}
	\hrule
	\vspace{-.2cm}
	\label{lst:interpreter}
\end{wrapfigure}
%
\begin{mathpar}
    \inferrule*[]
    {eval(E) \Rightarrow V,\; \sigma' = \sigma \ovl \{x \mapsto V\} }
    {\sigma \xrightarrow[x\;\bcmeq{}\;E]{} \sigma'}
\end{mathpar}
\vspace{-.1cm}


Concerning the performance aspect, we expect implementing an interpreter in Prolog to perform faster than translating to a B machine and then interpreting with \prob{}.
Still, for some verification and validation tasks, e.g. symbolic model checking, one could translate to B.
That approach would also enable code generation to other programming languages~\cite{b2program,b2program-mc,b2program-js-2} to achieve better performance.
Both aspects were analysed with Lustre~\cite{vumasterthesis} and may also apply to other formalisms.

\section{State Visualisation}\label{sec:visualisation}

Visualisation plays a crucial role in validating formal models, as it helps modellers and domain experts understand and check whether the behaviour is as expected.
Hence, \prob{} offers multiple ways to visualise state properties and interact with visualisations. 
This work improves the state visualisation of XTL models by an animation function and enables the use of the more advanced component \visb{}~\cite{visb}.

\vspace{-.2cm}
\paragraph{Animation Function.}

\begin{figure}[t]
	\centering
	\includegraphics[width=.9\linewidth]{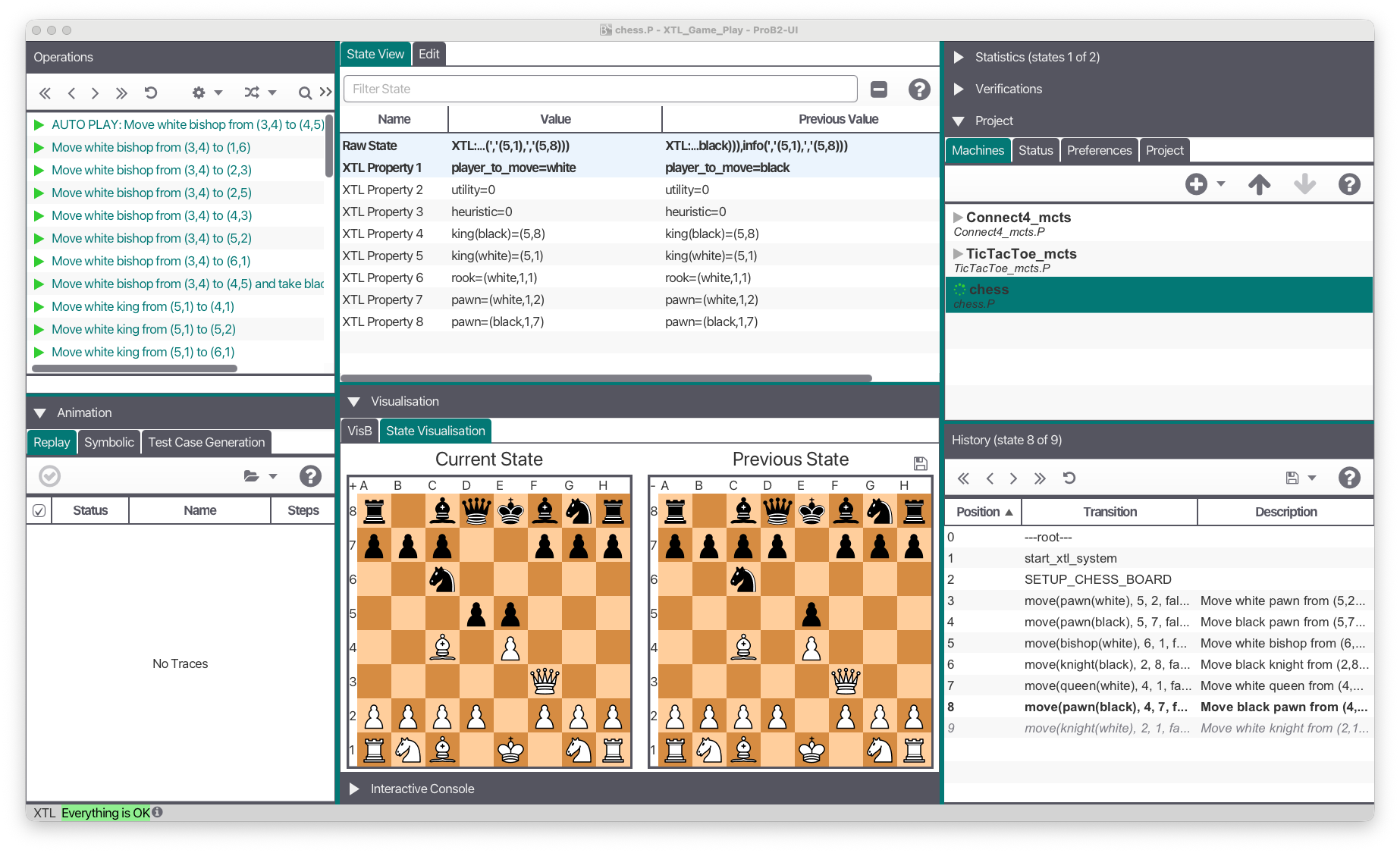}
	\vspace{-.3cm}
	\caption{\centering Interactive Chess Visualisation Defined by an Animation Function in \probtwoui{} \\ \tiny Chess Piece Icons by Cburnett via Wikimedia Commons, licensed under CC BY-SA 3.0 (\url{https://creativecommons.org/licenses/by-sa/3.0}).}
	\label{fig:chess-prob2ui}
\end{figure}

In XTL mode, it has already been possible to create a lightweight domain-specific state visualisation using special predicates targeting \prob{}'s ``animation function''.
This function defines a grid-based visualisation and transforms the current state into a matrix of coordinates, each mapped to an image or text.
The animation function is specified in Prolog using the predicate \texttt{animation\_function\_result(State,Matrix)}, where \texttt{Matrix} is a term of the form \texttt{((I,J),Img)}.
As part of this work, we integrate an HTML trace export into \probtwoui{}, which contains individual state visualisations for each trace step independently of \prob{} for inspection in a browser.


Our examples include an implementation of the chess game.
Allowed moves are encoded as transitions; several scenarios can be used as the initial state.
The current state of the chessboard is visualised using an animation function, as shown in \Cref{fig:chess-prob2ui}.
Images of the chess pieces are used as cell content (registered via \texttt{animation\_image(Nr,ImgPath)}).
Moreover, the visualisation is interactive by implementing right-click actions; the user can 
select from a list of available moves for the specific position (these actions are defined via \texttt{animation\_image\_right\_click\_transition(I,J,Act,State)}).
A complete HTML trace export can be found online~\cite{xtl-examples}.

\paragraph{VisB.}

Grid visualisations are especially useful for board games like chess, but they are limited when it comes to complex systems.
In this case, \visb{}~\cite{visb} provides a more flexible and customisable approach based on SVG graphics.
The user can provide a custom SVG, and the attributes of the SVG elements are updated according to the current state.
For example, this could be the colour of a circle representing a traffic light, as shown in \Cref{fig:tlprob2ui}.
As part of this work, we enabled the use of \visb{} for XTL specifications, allowing for more advanced 
visualisations.
We use the existing mechanism for loading B definition files containing \visb{} definitions, together with the specification file.
Using \visb{} definitions, one can specify SVG objects and corresponding attribute value updates.\footnote{A detailed summary of \visb{}'s syntax can be found at \url{https://prob.hhu.de/w/index.php?title=VisB}.}
For the traffic light in \Cref{fig:tlprob2ui}, the following definition registers an update of the \texttt{fill} attribute for the light with the SVG ID \texttt{cars\_red} depending on the state property \texttt{tl\_cars}:
\begin{minted}[ignorelexererrors=true]{prolog}
  VISB_SVG_UPDATES == rec(`id`: "cars_red", fill: IF STATE_PROPERTY("tl_cars") 
                          ∈ {"red", "red_yellow"} THEN "red" ELSE "black" END).
\end{minted}
That is, if the traffic light indicates \texttt{red} or \texttt{red\_yellow} in the current state, the element is filled with red, otherwise with black colour.
As \visb{} definitions are written in the B language, it is required to use the built-in function \texttt{STATE\_PROPERTY(Name)} to access the value of a state property. 
In this context, \texttt{Name} corresponds to a property of the special style \texttt{'='(Name,Value)} 
(a way to mimic identifiers, cf. \Cref{lst:traffic-light-example}).
As with the animation function, it is possible to register click events for SVG objects.
The following definition adds a click listener for the same element that triggers the event \texttt{switch\_tl\_cars} if the predicate is satisfied (the predicate can constrain the parameter values, here \textit{NL}, and the post state):
\begin{minted}[ignorelexererrors=true]{prolog}
        VISB_SVG_EVENTS == rec(`id`: "cars_red", event: "switch_tl_cars", 
                               predicate: "NL = STRING_TO_TERM(\"red\")").
\end{minted}
%
Our work introduces support for such event predicates, which are evaluated as B predicates.
However, a solution was required to use parameter values provided as Prolog terms in the B predicate, since they do not correspond to valid B data types.
We address this with a new built-in function \texttt{STRING\_TO\_TERM}, which wraps a B string and translates it to a 
data value of the type \texttt{term} for internal representation using basic Prolog functionality for reading terms from atoms.
Hence, for the above definition, a click triggers the transition \texttt{switch\_tl\_cars(red)} (if enabled).
Alternatively, one can provide the raw transition term directly, i.e. \texttt{event: "switch\_tl\_cars(red)"}.

\begin{figure}[t]
	\centering
	\begin{minipage}[b]{0.458\textwidth}
		\centering
		\includegraphics[width=\linewidth]{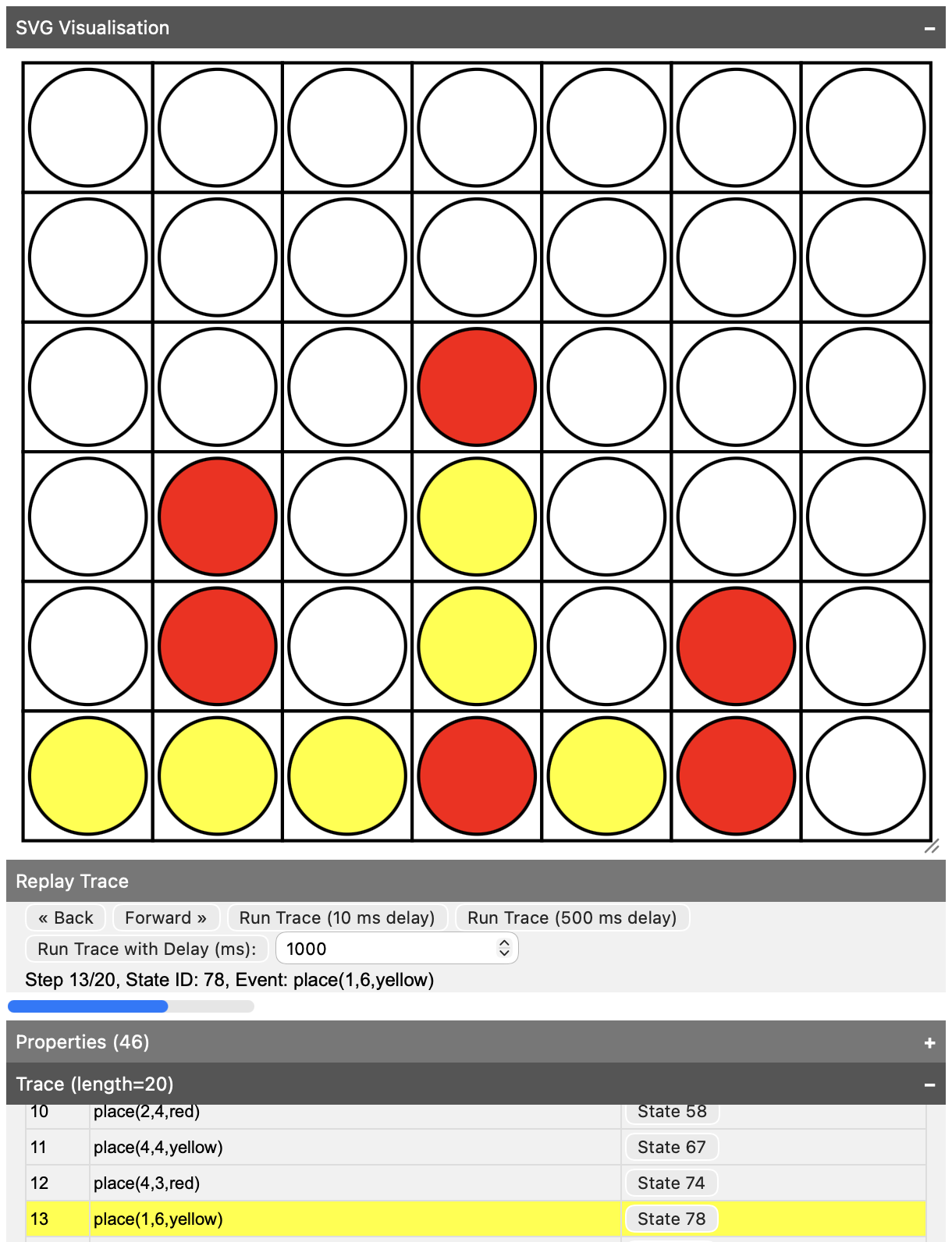}
		\caption{\visb{} HTML Export for Connect Four}
		\label{fig:visb-html}
	\end{minipage}
	\hfill
	\begin{minipage}[b]{0.535\textwidth}
		\centering
		\includegraphics[width=\linewidth]{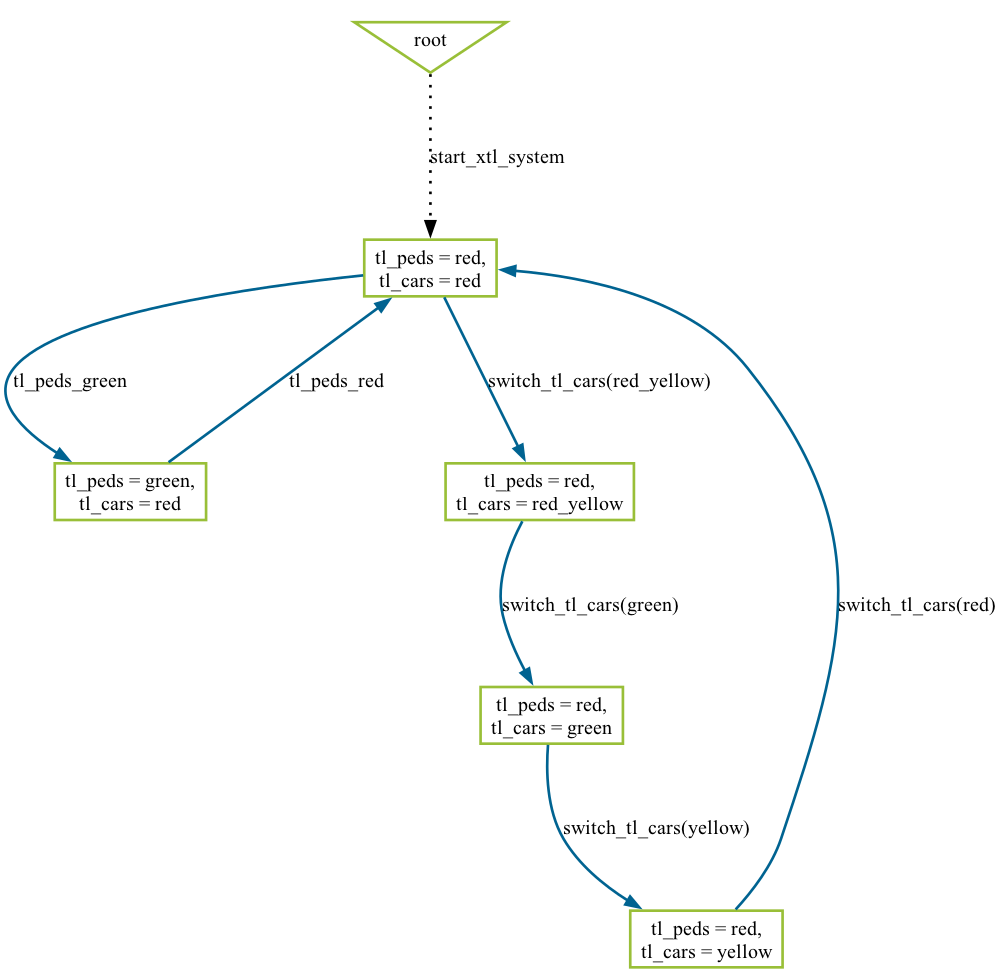}
		\caption{\centering State Space Visualisation\\for \Cref{lst:traffic-light-example} generated by \prob{}}
		\label{fig:tlstatespace}
	\end{minipage}
\end{figure}

Finally, the definition file must be linked in the XTL file using \texttt{prob\_pragma\_string("VISB\_DEFINITIONS\_FILE",Path)} or interactively in \probtwoui{}.

\visb{} offers an HTML trace export showing the visualisation for a state selected from the transitions.
The transitions can be controlled manually or played automatically with a certain delay.
The export has been made compatible with XTL models and extended by a table with the values of XTL state properties.
\Cref{fig:visb-html} shows a screenshot of a trace export for the Connect Four game (more details in \Cref{sec:games}).


\paragraph{Graph Visualisation.}

\prob{} offers additional graph visualisations with its interface to Graphviz~\cite{gansner2009drawing}.
A feature of interest when dealing with Prolog transition system is visualising the \emph{explored part} of the state space by plotting all the states and transitions between them.
\Cref{fig:tlstatespace} shows the state space of the traffic light system.
Other options include a graph showing the current trace and variations of the state space visualisation (e.g. highlighting the current state).

\section{New Animation Features}

This section presents the new features added to the animator, primarily to address requirements arising from the new \prob{} sequent prover based on the XTL mode~\cite{sequent-prover-iclp}.
A state of the sequent prover consists of a set of hypotheses and a goal, where all of them are predicates in first-order logic.
By applying proof rules that are encoded as XTL transitions, we aim to prove the goal based on the given hypotheses.
Additionally to the sequent prover, this section introduces a further example application, PCTL model checking, to highlight the applicability of the features in a different context.


\subsection{Transition Properties}


\prob{} allows one to store properties for transitions in the state space for B models.
We adapted this mechanism for XTL models so that one can add arbitrary Prolog terms as \emph{dynamic} transition properties (i.e. properties that can depend on parameter values and the state context in which the transition is available).
For this purpose, we supplement the transition predicate \texttt{trans/3} by \texttt{trans/4}, which accepts a list of transition properties as its last parameter.
The initial application includes user-friendly transition descriptions (for proof rule transitions), which is possible by adding a property \texttt{description/1}.
The descriptions are 
displayed in \prob{}'s GUIs, for example, at the top left of \Cref{fig:chess-prob2ui}, where chess move transitions are described by human-readable text.
The (PCTL) example below includes a second dynamic transition property in addition to \texttt{description/1}.

Another type of transition properties is \emph{static} properties, which do not depend on the state context of a transition, but provide general information about a transition. 
For this purpose, we add the new special predicate \texttt{trans\_prop/2}, which accepts a transition name and a property of this transition (one can provide any number of predicates with properties).
This predicate can be used, for example, to specify parameter names of a transition, e.g. \texttt{trans\_prop(t1,param\_names([p1,p2]))}.
Note that this restricts the arity of transition \texttt{t1} to the length of the list of parameter names (2) and makes the transition known to the \prob{} animator as disabled, even if no outgoing transitions are available.
This feature is useful for \emph{symbolic} transitions and trace replay, which we will describe in the following subsections.

\paragraph{Application: PCTL Model Checking.}

Recent work has integrated a model checker for Probabilistic Computation Tree Logic (PCTL) into \prob{} for probabilistic model checking.
With the new dynamic transition properties, probabilities can be easily assigned to transitions, for example: 
\begin{minted}{prolog}
    trans(tl_peds_green,lights(red,red),lights(green,red),
         [probability/0.6,description('switch pedestrian light to green')]).
\end{minted}
The probability could also be computed based on the state or parameter values.

\subsection{Symbolic Transitions}

When exploring a state, \prob{} computes all possible outgoing transitions.
However, some transitions require user input or have side effects that should not be triggered when searching for available transitions.
For example, this could involve the user-guided instantiation of a free identifier in a proof rule or the invocation of an external proof system.
Motivated by this, we introduced the concept of \emph{symbolic} 
transitions.
One can specify these transitions using the new predicate \texttt{symb\_trans/3} in the same way as \texttt{trans/3}, except that symbolic transitions are not automatically evaluated during animation and can be executed only using the ``execute by predicate''-mechanism of \prob{}.
The new predicate \texttt{symb\_trans\_enabled(TransitionName,State)} can be used to indicate whether a symbolic transition could \emph{potentially} be enabled. 

\paragraph{Execute Transitions by Predicate.} 
\begin{wrapfigure}{r}{0.46\textwidth}
	\centering
	\vspace{-.5cm}
	\includegraphics[width=\linewidth]{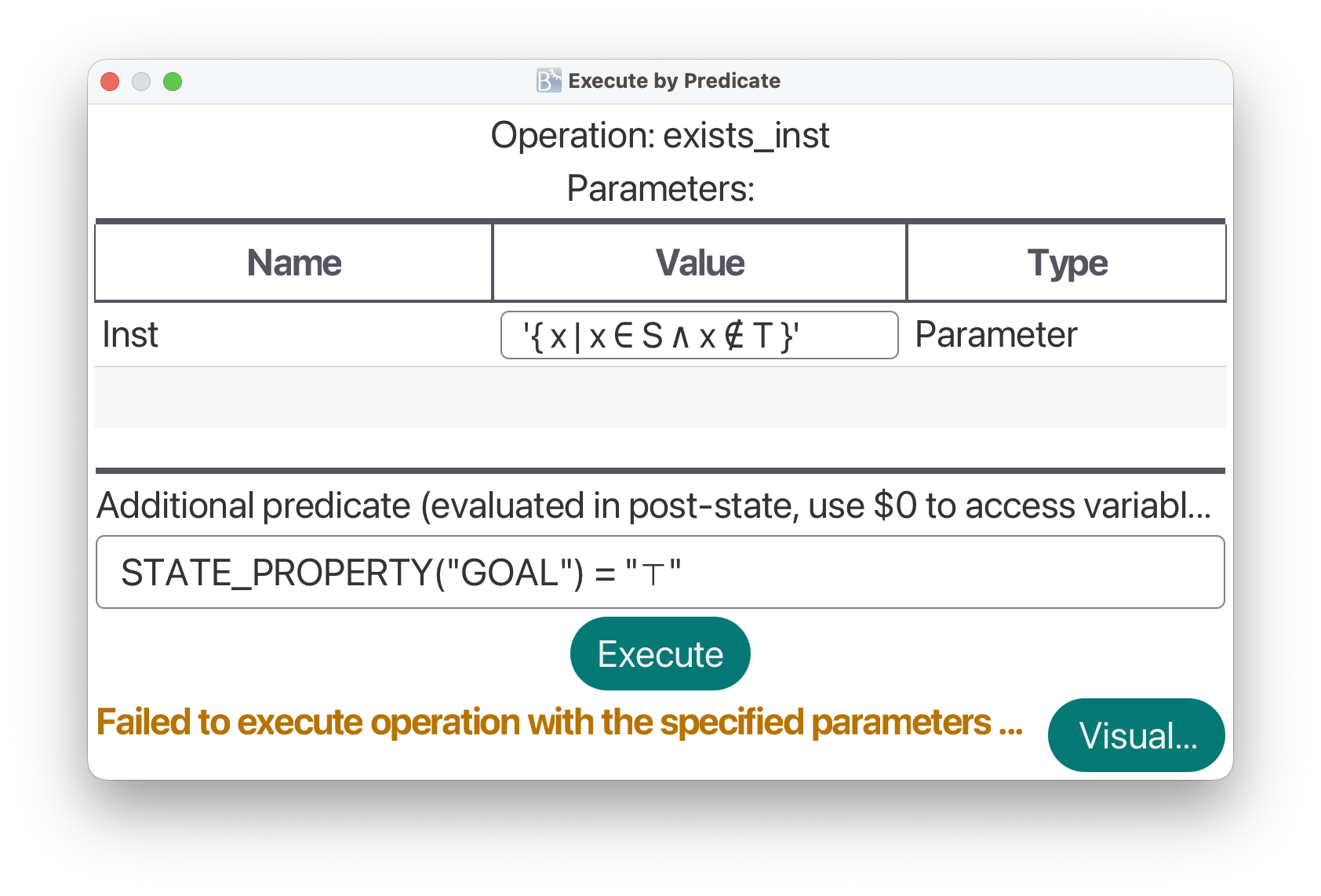}
	\vspace{-.7cm}
	\caption{Execute by Predicate in \probtwoui{}}
	\vspace{-.6cm}
	\label{fig:execpred-ui}
\end{wrapfigure}
This technique was previously available for B machines only and allows the user to find and compute possible transitions for the specified parameter values and further state properties. 
We implemented a variation for XTL, where the inputs are the current state term, the desired transition name, optional parameter values and an additional B predicate.
User input can be passed via the parameters or the additional predicate, as illustrated in \Cref{fig:execpred-ui}.
This information is translated into a transition term with the provided name as a functor and variables for the parameters.
The implementation unifies the parameters with the provided values during evaluation of the transition candidates.
If \prob{} finds a suitable candidate and the additional predicate is satisfied in the current state, it executes the corresponding transition.

\paragraph{Application: Proof Rules.}

The newly introduced feature is indispensable for the sequent prover, since advanced proofs often depend on user input.
It could be necessary to prove a goal that contains an existential quantifier.
In particular, the user cannot make progress by applying the rule alone, as a suitable choice for the instantiation is required to proceed with the goal.
The input (\texttt{Inst}) for the proof rule \texttt{exists\_inst} might look like in \Cref{fig:execpred-ui}.
The evaluation yields the transition term \mintinline{prolog}{exists_inst('{ x | x ∈ S ∧ x ∉ T }')}.
In this case, however, the predicate stating that the proof \texttt{GOAL} is true ($\top$) is not satisfied.
Therefore, \prob{} rejects the execution.

\subsection{Trace Replay}

After manual animation or automated simulation, the animator holds a certain sequence, i.e. a \emph{trace}, of transitions that have been performed to reach the current state.
We aim to save the trace as a separate file to use it as a regression test.
So far, a simplistic trace replay existed for the XTL mode, simply keeping the sequence of transition terms as a text file.
However, this can be ambiguous.
For instance, for each initial state, there is a \texttt{start\_xtl\_system} transition without parameters, which differs only in terms of the target state.
Since \prob{} does not store these states in the standard trace format, it selects the first available transition with a matching name during replay, which results in an imprecise replay.

\begin{wrapfigure}[5]{r}{0.61\textwidth}
	\vspace{-.4cm}
	\captionof{listing}{\centering Entry of a JSON Trace for an XTL Model}
	\vspace{-.25cm}
	\hrule
	\vspace{-.1cm}
	\small
	\begin{minted}{json}
{"name": "switch_tl_cars",
 "params": { "NL": "green" },
 "destState": { "xtl_state": "lights(red,green)." }}
	\end{minted}
    \vspace{-.15cm}
	\hrule
	\label{lst:json-trace-entry}
\end{wrapfigure}

To ensure correct trace replay, the JSON trace replay for B models has been reused and adapted for XTL.
As a result, we can use interactive trace replay~\cite{rodin-2025-interactive-trace-replay}, i.e. a user can reproduce traces step by step and replace transitions.
Each trace step corresponds to a JSON entry containing information such as the transition name, the parameter values by their name and the destination state (cf. \Cref{lst:json-trace-entry}).
For B models, \prob{} stores the destination state in the field \texttt{destState}, which contains a map of machine identifiers to their values in the corresponding state.
Since the concept of identifiers is not applicable in XTL mode, \prob{} now stores the raw destination state term with the special identifier \texttt{xtl\_state}.
For symbolic transitions, trace steps are replayed with the ``execute by predicate'' feature to unify user-provided values with the parameter variables.



\vspace{-.1cm}
\paragraph{Application: Replay of Proofs.}
The improved trace replay enhances the reproducibility of proofs with the sequent prover.
In some cases, a proof resembles an existing one, differing only in specific transitions 
which can be replaced during interactive trace replay.
On the other hand, if the user provides parameter values, the extended replay mechanism allows one to reload the corresponding trace without further interaction and manual adjustments by the user.

\section{Simulation}\label{sec:simulation}

\probtwoui{} offers a simulation feature for B models called \simb{}~\cite{simb}.
Its support has been extended for XTL, allowing \emph{simulation} of XTL transition systems.
This extension enables one to apply validation techniques such as Monte Carlo simulations with hypothesis testing and other statistical measures.

\vspace{-.15cm}
\begin{wrapfigure}[10]{l}{0.5\textwidth}
	\vspace{-.35cm}
	\captionof{listing}{\centering SimB Activation in JSON}
	\vspace{-.25cm}
	\hrule
	\vspace{-.1cm}
	\small
	\begin{minted}[linenos]{json}
{"id": "place",
 "execute": "place",
 "after": 10,
 "fixedVariables": {
    "player": "STRING_TO_TERM(\"yellow\")"
  },
 "transitionSelection":"uniform",
 "activating": ["auto_play_minimax"]}
	\end{minted}
	\vspace{-.15cm}
	\hrule
	\label{lst:json-simb}
\end{wrapfigure}
\paragraph{Timed Probabilistic Simulation.}
\simb{} introduces a concept to simulate formal models with timing and probabilistic behaviour, referred to as \emph{timed probabilistic simulation}.
The concept builds on \emph{activations} that execute an action after a specific time and subsequently trigger other activations.
In particular, one can specify how multiple activations trigger one another, or how to choose between multiple activations probabilistically.
\Cref{lst:json-simb} shows an example of an activation for Connect Four, which represents a random player.
The player places a yellow disc through the \texttt{place} event (after 10ms) at a random position, which is chosen uniformly (indicated by line 7).
Afterwards, the \texttt{auto\_play\_minimax} activation is triggered, which performs a move found by the \emph{Minimax strategy} (cf. \Cref{sec:games}).
Through the \emph{activating} attribute, \simb{} simulates the entire game automatically, with the random and Minimax player triggering one another.

\vspace{-.15cm}
\begin{wrapfigure}[5]{l}{0.45\textwidth}
	\vspace{-.35cm}
	\captionof{listing}{\centering SimB Listener in JSON}
	\vspace{-.25cm}
	\hrule
	\vspace{-.1cm}
	\small
	\begin{minted}[linenos]{json}
{"id": "player",
 "event": "place",
 "activating" : ["auto_play_minimax"]}
	\end{minted}
	\vspace{-.15cm}
	\hrule
	\label{lst:json-simb-interactive}
\end{wrapfigure}
\paragraph{Interactive Simulation.}

\emph{Interactive simulation}~\cite{simb-interactive} is an extended feature in \simb{} which allows one to trigger simulations interactively.
Technically, \simb{}'s activation concept is extended by \emph{listeners},
which are reactive elements that (1)~monitor whether an event is performed through animation, and (2)~trigger an activation afterwards.
With this feature, one can separate interactive and automatic components of a system and specify automatic processes triggered by a user event.
For example, users can play games such as Chess, Connect Four and Tic-Tac-Toe against automated strategies encoded in the XTL model: the user's move (see \texttt{place} in line 2 of \Cref{lst:json-simb-interactive}) triggers the autoplay transition (see line 3 of \Cref{lst:json-simb-interactive}), after which the listener waits for the next manual move.

\paragraph{Monte Carlo Simulation.}

\begin{wrapfigure}[10]{r}{0.44\textwidth}
    \centering
    \vspace{.1cm}
    \includegraphics[width=\linewidth]{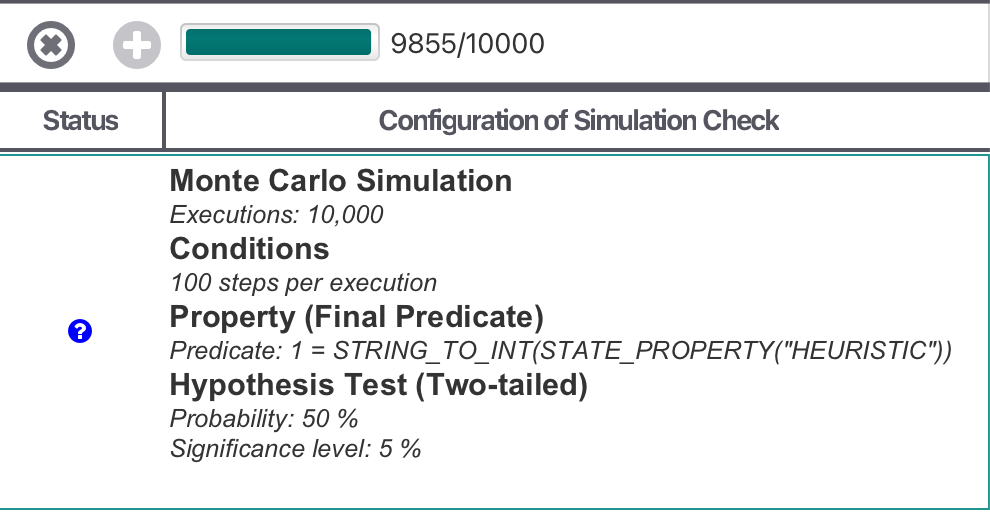}
    \vspace{-.5cm}
    \caption{Monte Carlo Simulation \probtwoui{}}
    \label{fig:simb-mc}
\end{wrapfigure}

Based on \simb{}'s automatic simulation features, one can perform Monte Carlo simulations with a specific number of execution runs.
One can specify an ending condition, specifying when each execution run stops.
Afterwards, one can formulate certain properties or expressions and use them for hypothesis testing and value estimation.
Due to the high number of runs in the Monte Carlo simulation, the user gains statistical confidence about the fulfilment of a property with a specific probability,
or that an estimated value is within a desired range.
\Cref{fig:simb-mc} shows a configuration for 10,000 runs, which evaluates a heuristic function in the final state and performs a two-tailed hypothesis test expecting that the predicate is satisfied with 50\%.
Traces from a Monte Carlo simulation can be saved using the new JSON trace export feature and then inspected individually.
In \Cref{sec:games}, we apply Monte Carlo simulations to evaluate strategies with the game Connect Four.

\vspace{-.17cm}
\section{Case Study: Game Play}
\label{sec:games}

In the following, we demonstrate the new simulation feature of \simb{} for XTL models.
Therefore, we discuss another use case where the logic of \emph{turn-based, deterministic, zero-sum games with perfect information}, such as chess, Connect Four and Tic-Tac-Toe, is encoded in XTL.
In addition to the game logic, the specification also contains the AI, implemented as Minimax search~\cite{neumann1928theorie}
with alpha-beta pruning~\cite{knuth1975analysis} and Monte Carlo Tree Search (MCTS)~\cite{coulom2006mcts,browne2012survey}.
More precisely, for each game, there are predicates for (1) playing one move manually, (2) playing one Minimax move (\texttt{auto\_play\_minimax}),
and (3) playing one MCTS move (\texttt{auto\_play}).
We use \simb{} to evaluate game strategies, in which a random player, a Minimax player, and an MCTS player play against each other.

When comparing different strategies, the focus is on determining the optimal move and teaching a computer to make that move. 
For this, a \emph{game tree} is constructed, with each node representing a player's decision, and each path from the root to a leaf signifying a potential game outcome.

With the \emph{Minimax} search strategy, we assume that our opponent will play the best possible move and then choose the best move for ourselves.
That is, we alternate between minimising and maximising levels of the game tree, computing the values for each node bottom-up.
For perfect play, one would have to expand the game tree until reaching a final state.
As this is not feasible due to the exponential growth, the search stops after reaching a certain depth.
Minimax uses a value (or heuristic) function to estimate the outcome.
Additionally, our Minimax implementation uses a random ordering to improve the performance of alpha-beta pruning.
Theoretical Minimax, when fully searched, ensures perfect play, allowing the first player to win always~\cite{neumann1928theorie,browne2012survey}.
%
%
 
\emph{Monte Carlo Tree Search} (MCTS) is an alternative approach that searches the game tree by executing simulations from the current state until a player wins.
Initially, MCTS chooses actions randomly, but as the process continues, data on node visitation frequencies and win rates refine the choice of actions, leading to progressively less randomness.
With an increasing number of simulations, MCTS approaches optimal play and achieves perfect results with an infinite number of simulations~\cite{bowen1999animating}.
 
 
The \prob{} examples~\cite{prob-examples} contain generic implementations of Minimax and MCTS that can be included into any XTL game specification, provided that the necessary predicates for the value function and game moves are available.
As a particular example and a classic game for AI benchmarks~\cite{sym18020293}, we examine Connect Four.
In Connect Four, two players take turns placing discs in the columns.
The aim is to be the first player to get four discs in a horizontal, vertical, or diagonal row.
The game is solved~\cite{allis1988knowledge}, meaning the outcome is predictable from any state with perfect play.
The first player can force a win by starting in the middle column; other statements can be made if they choose one of the adjacent columns. 

For evaluation, we consider random play, Minimax with a search depth of two moves and MCTS (without any time limit per move).
We conducted 10,000 Monte Carlo simulations (i.e. matches) for each combination of the three strategies, once with the first move chosen freely and once with it fixed to the middle column.
We then evaluated the probability of the first player winning.
One could either perform a hypothesis test to validate the expected probability, or use an estimator to estimate the average final value of the heuristic function (ranging from -1 for loss to 1 for win).
\Cref{fig:simb-mc} shows an example configuration.
We encode the \simb{} activations as described in \Cref{sec:simulation}.
In addition, we evaluate the average trace length (i.e. the number of moves, top right) and the performance in terms of execution time (bottom right).
The results are presented in \Cref{tbl:simulation}.

\begin{table}[t]
    \centering
    \caption{\centering Simulation Results for Connect Four (10,000 or $\dagger$1,000 runs) \\ \footnotesize Percentage of Wins by the First Player (1), Average Trace Length and Total Duration of the Simulations}
    \label{tbl:simulation}
        \setlength\tabcolsep{8pt}
        \begin{tabularx}{\textwidth}{r | l l l | l l l}
            \toprule
            \multicolumn{1}{c}{\footnotesize First Move} & \multicolumn{3}{c}{Free Choice} & \multicolumn{3}{c}{Fixed {\footnotesize (Middle Column)}} \\
            ~[\% {\scriptsize \makecell{length\\min.}}]\hspace{.2cm} \diagbox{\bf 2}{\bf 1} & \textbf{Random} & \textbf{Minimax} & \textbf{MCTS} & \textbf{Random} & \textbf{Minimax} & \textbf{MCTS} \\
            \midrule
            {\bf Random} & 55.7 {\scriptsize \makecell{22.3\\13.1}} & 96.0 {\scriptsize \makecell{18.0\\21.9}} & 100.0 {\scriptsize \makecell{9.7\\266.3}} & 64.4 {\scriptsize \makecell{21.8\\11.2}} & 97.3 {\scriptsize \makecell{16.3\\13.4}} & 100.0 {\scriptsize \makecell{9.7\\214.9}} \\
            {\bf Minimax} & 7.3 {\scriptsize \makecell{20.3\\18.3}} & 46.8  {\scriptsize \makecell{30.7\\20.6}} & 98.9 {\scriptsize \makecell{19.5\\511.0}}  & 9.9 {\scriptsize \makecell{21.4\\19.7}} & 52.0 {\scriptsize \makecell{29.8\\19.7}} & 98.8 {\scriptsize \makecell{19.8\\443.5}} \\
            {\bf MCTS} & 0.0 {\scriptsize \makecell{12.5\\373.9}} & 0.5 {\scriptsize \makecell{24.5\\730.0}} & $\dagger$ 44.2 {\scriptsize \makecell{32.8\\147.3}} & 0.0 {\scriptsize \makecell{14.6\\403.6}} & 1.1 {\scriptsize \makecell{25.6\\718.8}} & $\dagger$ 47.1 {\scriptsize \makecell{32.9\\153.8}} \\
            \bottomrule
        \end{tabularx}
        \vspace{-0.2cm}
\end{table}
 

Obviously, a random player has almost no chance against Minimax and no chance at all against MCTS.
The weakness of Minimax is likely due to the shallow search depth we chose.
Taylor and Stella~\cite{taylor2024evolutionaryframeworkconnect4testbed} come to a similar conclusion for Minimax with depth three.
This result also becomes evident when comparing Minimax against MCTS, where MCTS won around 99\% of the games.
However, this result was expected as outlined by Sheoran et al.~\cite{sheoran2022solving}, who compared an optimised Minimax algorithm with MCTS.
The results of the comparison between random and MCTS are also the same (100\%). 
They claim that with increasing search depth (four and five), Minimax can beat MCTS in more than 60\%.
Using a sample of 100 games, we could indeed show that the probability for Minimax wins increases from a depth of five onwards, although performance decreases.
Interestingly, the results show that the first player does indeed have an advantage in a random vs. random match.
However, further research is needed to determine why this is not the case with Minimax and MCTS.
As expected, the probability of winning increases for all strategies if we restrict the first move to the middle column.

The performance is satisfactory for random and Minimax (partly due to the small search depth).
It was possible to simulate 10,000 matches in less than 20 minutes without any issues.
MCTS, on the other hand, slows down significantly, resulting in run times of several hours.
Depending on the number of remaining moves, the computation times per move can reach up to 500 ms.

An interesting finding is that MCTS only requires an average of 9.7 moves to beat the random player, whereas Minimax requires 18.0 and a random opponent 22.3 moves.


One can perform similar evaluations on the other implemented games.
For games with a manageable state space, model checking is another exploration method.
For instance, \prob{} model checks the Tic-Tac-Toe model in approximately 200 ms, analysing 5478 states and 16168 transitions.
Our model is available online together with the other games~\cite{xtl-examples}. 




\section{Practical Applications and Related Work}

The XTL mode has been used for teaching in several iterations of our university courses covering both introductory and advanced topics on logic programming.
The modelling of small systems is taught to first-year Prolog students.
With the XTL mode, it is easy to switch from pure code to an experimental mode where students can repeatedly test their own implementations.
Complementary visualisations can help students better understand the system’s state.
Unlike approaches using interactive documents, such as Jupyter notebooks~\cite{brecklinghaus2022jupyter,sartor2025teaching} or active learning documents~\cite{morales2023teaching}, the XTL mode enables exploration of a Prolog model within a formal verification tool.
Similar to our case study, Krings and Körner~\cite{krings2019prototyping} propose to formalise the rules of games in the context of teaching formal methods.
Our advanced course on logic programming involves a project in which the students develop an interpreter in Prolog.
The XTL mode is particularly well suited to this task because semantic rules can be easily implemented and immediately tested (cf. \Cref{sec:background}).
In recent years, interpreters have been implemented for subsets of Java bytecode, Petri nets and WebAssembly.

There exist further applications, such as for model checking of security protocols~\cite{vellathesis}.
More examples are available in the \prob{} examples collection~\cite{prob-examples}.
Sterling et al.~\cite{sterling1996animation} implemented a compiler from Z to readable Prolog code for animating formal specifications.
In another related work, a Prolog-based animator for the Verilog Hardware Description Language has been developed~\cite{bowen1999animating}. 
Körner et al.~\cite{körner2020performancebytecodeinterpretersprolog} discussed the performance of bytecode interpreters implemented in Prolog.

\textsc{Clingraph}~\cite{clingraph} is a visualisation tool for \emph{answer set programming} (ASP) using Graphviz~\cite{gansner2009drawing}. 
In particular, \textsc{Clingraph} supports SVG visualisation of the computed solutions, dynamic problems (whose solutions are similar to a trace in formal methods), and the program structure.
Another visualisation tool is \textsc{ASPVIZ}~\cite{aspviz} which allows one to present ASP solutions from a domain-specific perspective.
Using \textsc{ASPVIZ}, one can also visualise the solutions as an animation (in the context of computer graphics, i.e., a sequence of frames).
Bertagnon and Gavanelli present \textsc{ASPECT}~\cite{10.1093/logcom/exae042}, which is a sub-language of ASP that can be used to present the solver's output graphically.
Using \textsc{ASPECT}, the results can be converted to \LaTeX{} to produce vector graphics.
Another tool to visualise and debug Prolog programs is PrettyCLP~\cite{stalla2011prettyclp}. 
Its visualisation features include SLD trees and and CLPFD constructs.

\prob{} also supports domain-specific visualisations of the underlying system based on SVG graphics.
The main difference is that \prob{} treats XTL programs as a state-based formalism, the other tools focus on the presentation of the ASP solutions.

\vspace{-.1cm}
\section{Conclusion}

In this work, we presented the capabilities of \prob{}'s XTL mode for animating transition systems specified in Prolog.
We demonstrated new features, in particular, improved state visualisation and simulation applied to a detailed study of strategies with the Connect Four game.
We argue that XTL provides a simple and flexible interface for validating and verifying Prolog transition systems, especially suited for writing simple interpreters to quickly connect other (formal) languages to the \prob{} tooling.  
%
%
One key application involves the sequent prover, which we intend to enhance in the future using an iterative deepening approach, similar to the games involving automatic proof strategies.

With our new simulation for XTL models and Monte Carlo simulations in particular, we can validate Prolog specifications and obtain indications of potential weaknesses in our implementation, e.g., Minimax losing against a random player in a few scenarios.
%
In the future, we plan to evaluate game strategies using AI-driven simulations via \simb{}'s interface to external simulations~\cite{validation-rl-safety-shields}, such as a reinforcement learning agent that controls the simulation.


\bibliographystyle{eptcs}
\bibliography{references}
\end{document}